\documentstyle[12pt,epsf]{article}
\begin{document}

\include{psfig}
\title{Mechanical Properties of a Model of Attractive Colloidal Solutions}
\author{
E. Zaccarelli$^1$, G.Foffi$^1$, K. A. Dawson$^1$, F. Sciortino$^2$ and 
P. Tartaglia$^2$\\
{\small $^1$ Irish Centre for Colloid Science and Biomaterials, Department of
Chemistry,} \\{\small University College Dublin, Belfield, Dublin 4, Ireland}
\\ $^2$ {\small Dipartimento di Fisica, Universit\`{a} di Roma La Sapienza}\\
{\small and Istituto Nazionale di Fisica della Materia, Unit\`{a} di Roma La
Sapienza}\\ {\small Piazzale Aldo Moro 2, 00185 Roma, Italy} }

\date{\today}
\maketitle

\begin{abstract}
We review the nature of glass transitions and the glasses arising from
a square-well potential with a narrow and deep well. Our discussion is
based on the Mode Coupling Theory, and the predictions of glasses that
we make refer to the `ideal' glasses predicted by that theory. We
believe that the square well system well represents colloidal
particles with attractive interactions produced by grafted polymers,
or depletion interactions.  It has been recently shown that two types
of glass, an attractive and a repulsive one, are predicted by MCT for
this model. The former can form at quite low densities.

Most of our attention is directed at the mechanical properties of the
glasses predicted by this theory.  In particular we calculate the
elastic shear modulus at zero frequency and the longitudinal stress
modulus in the long wavelength limit. Results for both are presented
along the glass-liquid transition curves, and their interesting
behaviour is explained in terms of the underlying physics of the
system.

\end{abstract}



\section{Introduction}
In recent years the idea that colloidal particles can form glassy
structures has been established in a number of very interesting
experimental and theoretical works. Most attention has been focused on
colloidal particle systems that are dominated by repulsive
interactions, for which, at high packing fraction values, the glass
represents an alternative packing to the usually more favourable
crystal structure. These systems are well represented by a simple hard
sphere model \cite{hansen}.  Where repulsive interactions dominate,
the loss of ergodicity is lost  due  
to blocking of the
movements of particles by the quite dense surrounding cages formed by
their nearest neighbours.  In colloidal systems, Mode Coupling Theory
\cite{gotze} has played a leading role, interpreting and rationalizing
some of the observations \cite{bengt,leuth}, and achieving quite
acceptable numerical agreement in comparison to experiments
\cite{vanmegen,vanmegen2,vanmegen3}.

Recently a new type of glass has arisen as the focus of interest
\cite{fuchs,sciotar,fuchs2,fuchs3,statphys,us}.  This has been called
the attractive glass. By attractive glass we imply that the loss of
ergodicity is driven largely by strong short-ranged attractive
interactions, in other words the `stickiness' of the particles to each
other eventually dominates the thermal motions, and the system
freezes. Thus, close packing is no longer necessary for a glass to be
stable, and it transpires that such glasses can form at densities much
lower than close packing.

Once having established the distinct nature of the attractive glass
phenomenon, it is unsurprising that new phenomena should be associated
with the system.  For instance, MCT predicts that, for well defined
values of the temperature or packing fraction, the repulsive and
attractive glasses differ only by their dynamical properties (as
opposed to structural differences).  Many questions remained
unanswered in this arena, and more research, using different
approaches must be developed before one has great confidence in the
conclusions drawn so far. Nevertheless, the basic results emerging
from MCT seem reasonable, and are reproducible, using a variety of
different input static structure factors \cite{us}.

The situation in experimental studies is even less clear. It may be
that colloidal glasses driven by attractive interactions have been
observed and studied for some time, without clear recognition of their
distinctive nature. Certainly some of the reported kinetically
arrested states in colloidal systems, usually identified as `gels',
appear to arise at densities considerably lower than close packing
\cite{verduin,ilett,verhaegh}. Given that such observations appear to
be associated with quite strong short-ranged attractions driven by
depletion forces, or grafted chains onto larger particles, it would
seem likely that indeed such systems are examples of those we
currently discuss.  On the other hand, as yet, there appear to be no
clear reports of colloidal glass-glass transition in the experimental
literature, but the typical logarithmic decay of density correlations,
that MCT predicts to happen close to an $A_3$ singularity \cite{sjogren}
has been observed in the past, for the glass transition
of some polymeric systems \cite{scott,nakajima}. More recently a
logarithmic decay has been reported in a micellar system with
short-ranged attractive interactions, and this may be related to a
proximate glass-glass transitions as these authors point
out \cite{mallamace}. The clear recognition of two different glasses
would be one clear and unambiguous signal that a distinctive
attractive glassy state has been observed.  Since this is expected to
be an issue of considerable interest to experimentalists, it will be
important to discuss those measurements that could differentiate the
glasses, and this is one aim of the present work.
        
In this paper it is our intention to discuss the mechanical properties
of the glasses formed when repulsive and attractive interactions are
present, and where, under some conditions, the latter may become
dominant.  We shall work within the framework of the ideal MCT, and
present results for the linear elastic shear modulus at zero frequency
, $G'(\omega=0)$, and for the longitudinal stress modulus in the
hydrodynamic limit, corresponding to zero frequency and long
wavelengths, $m_0$, to illustrate our point of view.  The study of the
elastic shear modulus has previously been addressed in \cite{fuchs},
but there a different potential is used.

Our reasoning for presenting these results has two underlying strands
of rationale. Firstly, we have outlined the interesting and novel
features we believe to be present for colloidal systems where there
are strong short-ranged potentials. However, we note that experiments
in dense colloidal systems are far from simple, and the number of
techniques that can be reliably applied to examine these questions are
much smaller than in dilute solutions due to multiple scattering and
other limitations implied at high concentrations.  Furthermore, when
one works on the boundary between liquids and amorphous solids there
are further restrictions in the options for experimental techniques
that may be applied.  In fact, the mechanical responses of an
amorphous materials are amongst the most simple and reliable methods
of characterisation. Even the bulk shear and longitudinal moduli are
quite helpful in forming an initial assessment of the material itself,
and of course they are themselves amongst the simplest parameters
characterising the transition from a liquid to a glass, and as we
shall see, even between different glasses.  Combined with this, it is
clear that most practical applications of amorphous materials composed
from colloidal particles will involve a strong interest in mechanical
properties of the glass. Having said all this, the reader should be
aware that many attractive glasses may be quite fragile, and some
thought will have to be applied to finding systems and methods where
measurements can be made.  In any case, these comments comprise the
first strand of reasoning for our interest in the mechanical moduli of
attractive glasses.
       
The second reason for our interest is quite different, and involves a
deeper analysis of the basis of the theory that we use to describe
glassy systems. That is, rarely is MCT applied to a system simple
enough that many issues can be worked out in detail, but at the same
time that there be a complete knowledge of the `phase diagram' and its
dynamics. In our particular case, we shall study the square-well
potential model, with very short-ranged attractions. Some of the basic
predictions for structure and dynamics of this model have been worked
out in detail within the MCT formalism \cite{us}.  We have observed
that the attractive glass introduces a new richness into the study,
and it naturally becomes of interest to understand what presumptions
MCT is making about such phenomena. We shall be interested in such
questions as, what part of the physics drives the attractive glass
transition within MCT, and whether this understanding can shed some
light on the nature of the theory itself.  Finally, as an aside we
shall also seek to understand which physical characteristics of the
colloidal particle and of the whole system sets the scale for the
mechanical properties of a colloidal glass. Again, if MCT is correct,
this will be of considerable practical interest.
        
In the light of these comments it is now possible to proceed to
introduce the fundamental equations that define MCT, and the
mechanical responses that we calculate from it.

\section{Theory}

Mode Coupling Theory provides a description of the structural
relaxation of super-cooled liquids. The variables of interest are the
normalized density correlators defined as,
\begin{equation}
\phi_q(t)=\langle \rho^*_{\bf q}(0)\rho_{\bf q}(t)\rangle/
\langle |\rho_{\bf q}|^2\rangle 
\end{equation}
where $\rho_{\bf q}(t)=\sum_{j=1}^N e^{i{\bf q}\cdot {\bf r}_j(t)}$,
with $N$ being the number of particles in the system.  By using the
Mori-Zwanzig formalism, it can be shown \cite{gotze} that the
equations of motion for the variables $\phi_q(t)$ are,
\begin{equation}
\label{mctdyn}
\ddot{\phi_{q}}(t)+\Omega_{q}^{2}\phi_{q}(t)+\nu_{q}\dot{\phi_{q}}(t)+
\Omega_{q}^{2}\int_{0}^{t}m_{q}(t-t')\dot{\phi_{q}}(t')dt'=0
\end{equation}
for Newtonian dynamics. The two quantities $\Omega_{q}$ and $\nu_{q}$
are respectively the characteristic frequency of the phonon-type
motions of the fluid, and a term that describes instantaneous damping,
the latter arising from the `fast' contribution to the memory
function. They are defined as,
\begin{equation}\Omega_{q}=\frac{q^{2}k_{B}T}{mS(q)} \end{equation}
\begin{equation}
\nu_{q}=\nu_{1}q^{2} 
\end{equation}
and typically one chooses $\nu_{1}=1$ in the calculations. Equations
(\ref{mctdyn}) are formally exact for a set of N particles.\\ To
describe the dynamics of colloidal suspensions, they have been
modified neglecting the inertia term and including the solvent
contributions \cite{franosch}.  Thus, we have
\begin{equation}
\dot{\phi_{q}}(t)+q^2D_q^s\left\{\phi_q(t)+\int_{0}^{t}m_{q}(t-t')
\dot{\phi_{q}}(t')dt'\right\}=0
\label{mctbrownian}
\end{equation}
where $D_q^s$ is the Brownian short-time diffusion.\\ The crucial
approximation of MCT consists of giving an explicit factorized form
for the memory kernel in eqs. (\ref{mctdyn}) and (\ref{mctbrownian}) as,
\begin{equation}
\label{memory}
m_{q}=\frac{1}{2}\int \frac{d^{3}k}{(2\pi)^{3}}V( {\bf q},{\bf
k})\phi_{k}(t)
\phi_{|{\bf q}-{\bf k}|}(t)
\end{equation}
and the  vertex functions are the coupling constants of the theory,
\begin{equation}
\label{vertex}
V( {\bf q},{\bf k}) = \frac{\rho}{ q^{4}}\left({\bf q}
\cdot( {\bf q} - {\bf k}) 
c_{|{\bf q}-{\bf k}|}+ {\bf q}\cdot {\bf k} c_{k}\right)
^{2}
S_{q}S_{k}S_{|{\bf q}-{\bf k}|}
\end{equation}

In the static limit $t\rightarrow \infty$, independently on the type
of microscopic dynamics, the density correlators $\phi_q(t)$ tend to a
finite value $f_q=\langle \rho^*_q(0)\rho_q(\infty)\rangle/ \langle
|\rho_q(0)|^2\rangle$, known as the non-ergodicity factor, if the
system is kinetically arrested. This loss of ergodicity for the
density correlators is seen as the transition to a kinetic glassy
state within MCT.  Thus, the equations (\ref{mctdyn}) become in the
static limit,
\begin{equation}
\frac{f_{q}}{1- f_{q}} = \frac{1}{2} \int \! \! \frac{d^{3}k}{(2\pi)^{3}} \,\,
 V( {\bf q},{\bf k})f_{k}f_{|{\bf q}-{\bf k}|}  
\label{mctstat}
\end{equation}
It is clear that $f_q=0$ always corresponds to a solution of equations
(\ref{mctstat}).  This corresponds to an ergodic state of the system
in which the correlations decay for long time.  For some critical
values of the thermodynamic parameters(control parameters), such as
temperature and density, there appear bifurcations of the solutions of
equations (\ref{mctstat}), that produce non-zero solutions.  These
correspond to the non-ergodic states of the system, and given that
there is no positional order in the system we identify these solutions
as glasses.  The bifurcations can be multiple, up to the number of
control parameters of the system. Thus, when a bifurcation gives rise
to more than two solutions of equations (\ref{mctstat}), there will
exist multiple solutions with finite non-ergodicity factors. In this
case, MCT predicts that only the state corresponding to the largest
value of $f_q$ is the long-time limit solution of the
equations \cite{gotze,gotze2}.

With an input structure factor, we can now solve the MCT equations for
the non-ergodicity parameter, and it is possible to calculate the
`phase diagram' of the system, localizing the regions in the
thermodynamic parameter space where the system is in the fluid
($f_q=0$) or in the glassy state ($f_q>0$), and also some mechanical
properties of the glass itself.  Note that by `phase diagram' we mean
here that the fluid and glassy states of the system are identified,
the latter being non-equilibrium states of matter.

A particularly interesting physical quantity is the elastic shear
modulus, $G'(\omega)$.  From the shear viscosity for a colloidal
system \cite{bengt,visco}, it is possible to evaluate the
elastic shear modulus, within the MCT approximation, in the limit
$\omega \rightarrow 0$, to give \cite{fuchs},
\begin{equation}
\label{G'}
G'(\phi,T)={{d^3}\over{60\pi^2}}{\int_0^\infty}dk k^4
{\left({{dlnS_k}\over{dk}}f_k \right)}^2
\end{equation}
where $G'$ is in units of $(k_B T)/d^3$, where $T$ is the 
temperature of the system and $d$ is the diameter of a particle.

Another property that can be examined by experimentalists is the
longitudinal stress modulus $m_0$. We will discuss only the
hydrodynamic approximation of this quantity \cite{bengt,mayr}, that is
easily obtained taking the limits $q\rightarrow 0$ and $t\rightarrow
\infty$ in the equations (\ref{memory}) and (\ref{vertex}), giving
that,
\begin{equation}
\label{m0}
m_{q=0}(\omega=0)=m_0=\int_0^\infty V_k f_k^2 dk
\end{equation}
where the long wavelength limit of the vertex function is given by,
\begin{equation}
\label{V0}
V_k=\rho S_{q=0} \left( \frac{S_k k}{2 \pi} \right)^2 \left[ c_k^2+\frac{2}{3}
\left(k \frac{\partial c_k}{\partial k}\right) c_k+\frac{1}{5}
\left(k \frac{\partial c_k}{\partial k}\right)^2 \right]
\end{equation}

It is possible to relate $m_0$ to the velocity of sound in the glass
(solid) compared to that in the liquid, at the transition line. Thus,
we observe that the speed of sound in a liquid is given in terms of
the compressibility of the liquid whilst the formation of a solid
leads to a finite memory kernel at long times, and consequently an
increment, $m_0$:
\begin{equation}
\label{sound}
\frac{v_\infty}{v_0}=\sqrt{1+m_0}
\end{equation}
where $v_0$ and $v_\infty$ are respectively the limiting low and
high-frequency sound speeds.

 Since the longitudinal modulus is a close
equivalent of the shear modulus, but for extensional distortions, it
is natural that we should seek to calculate and compare the two
quantities. We would like to point out to experimentalists who may be
interested in this field that the theory, as applied to ratios of
quantities at the transition may be quite well modeled by the theory,
and less dependent on model details. We therefore recommend some
thought on how experiments of this type might be carried out.  


We have solved \cite{us} the MCT equations for the ideal
glass-transitions (\ref{mctstat}) for a system of mono-disperse
colloidal particles, interacting via a square well potential, with a
very narrow-range attractive well, defined as

\begin{equation}
\label{potential}
V(r)=\left \{\begin{array}{cc} \infty \ \ \ \ \ r < d\\ -\beta u_0 \ \
\ \ \ \ \ d <r < d+\Delta
\\ 0  \ \ \ \ \ \  d+\Delta < r \end{array} \right.
\end{equation}
where $\beta = \left({k_B T}\right)^{-1}$, with  $k_B$ the Boltzmann's
constant, and $\Delta$ is the well-width, which can be related to the small 
parameter $\epsilon=\Delta/(d+\Delta)$.

To calculate numerically the equilibrium structure factors $S(q)$ of
such a system, one can use the Mean-Spherical Approximation (MSA),
Percus-Yevick, or other closure relations for the Ornstein-Zernike
integral equation \cite{us}.  Ref.\cite{us} compares many of the
important features of the statics and dynamics of these systems using
MSA and Percus-Yevick (PY) approximations.  In view of the good
agreements found between these closures, we shall here present only
calculations based on the PY closure.  The phase diagram for different
values of the well-width of the potential has been presented in
Ref. \cite{us}.  In particular, for the mechanical results, we will
focus our attention on the case where $\epsilon=0.03$.  Here we report
in Figure \ref{fig1} more details and present both the glass
transitions as well as the liquid-gas spinodal, which may be regarded
as an approximation to the equilibrium phase diagram. This provides
the reader with an overview of where we expect to find all of the
phenomena discussed later.  Thus, in Fig. \ref{fig1} we can see the
curves labelled respectively $B_1$ representing the fluid-repulsive
glass, and $B_2$ representing the fluid-attractive glass
transition. In the inset, we have shown in more detail the
attractive-repulsive glass transition curve with its end-point,
labelled as $A_3$. Beyond this point, the two types of glasses become
indistinguishable.  Also, the dashed curve in the low volume fraction
region represents the gas-liquid spinodal, calculated via an expansion
in $\epsilon$ around the Baxter potential \cite{chen} and in good
agreement with numerical calculations for the square-well potential,
at least to the right-hand side of the critical point (larger volume
fractions). We comment here that the left-hand side branch of the
spinodal is singular for a Baxter model, because it corresponds to
complex values of the characteristic parameters $\lambda$, and also
for such low packing fractions and temperatures the numerical PY
solution of the square-well potential was not reliable. Thus, the only
true meaningful branch of the spinodal in Fig. \ref{fig1} is the
`liquid' branch, that corresponds approximately to $\phi \geq 0.14$,
where we estimated that the critical point is located, corresponding
to a critical temperature of approximately $0.3$.

It is interesting to note that, whilst the curve $B_1$ is almost
vertical, $B_2$ is largely horizontal and the two meet forming a
non-zero angle.  The fact that the repulsive glass-liquid transition
curve is vertical is ensured by the fact that glassification is driven
only by the hard core, which lacks any energy scale. On the contrary,
the attractive glass-liquid one, being fairly horizontal, implies that
there is a single well-characterized energy scale that drives the
glassification.

\section{Results}

\begin{figure}
\centerline {\epsfxsize=17.0cm
             \epsfbox{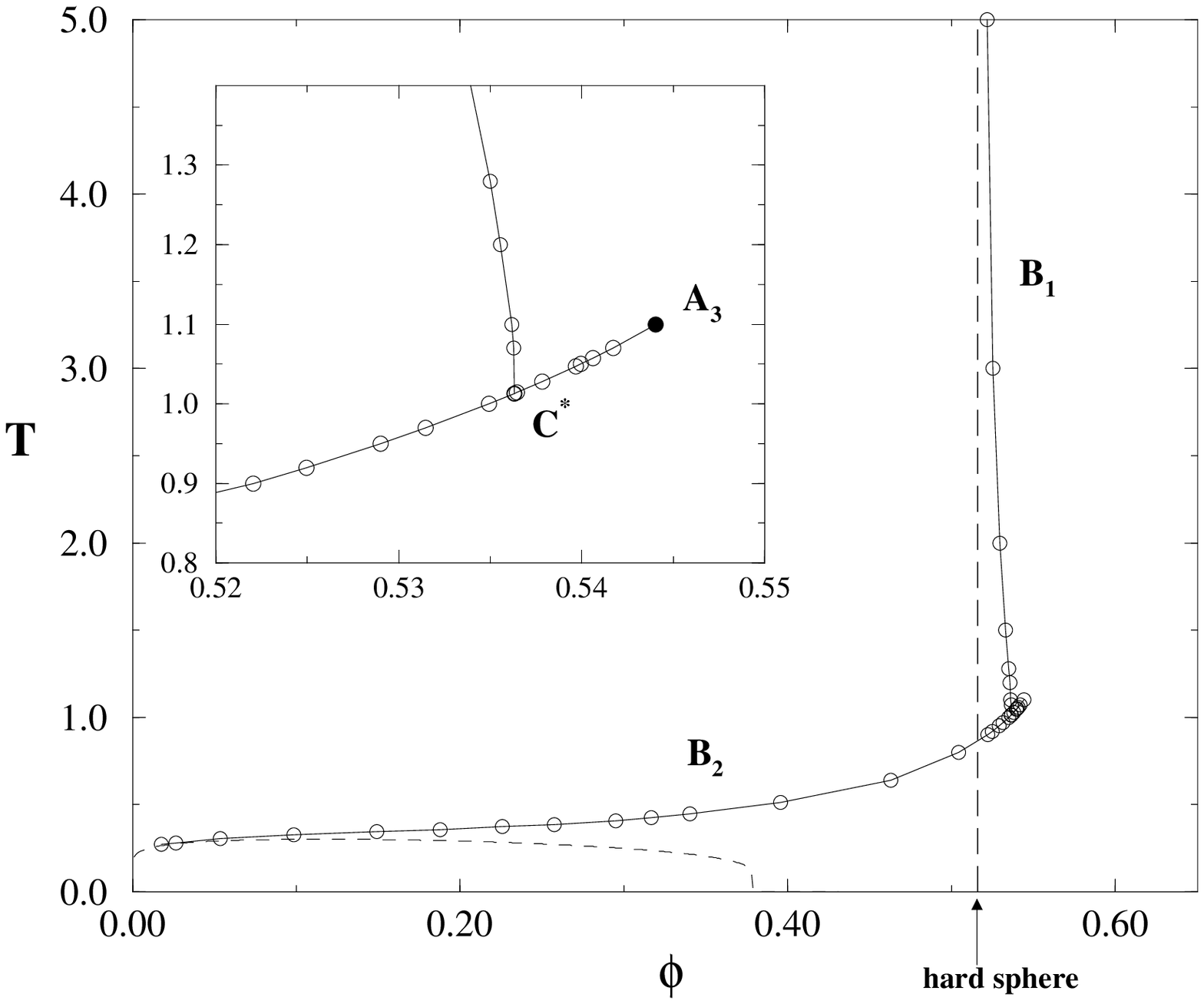}}  
\caption{
Phase diagram of a colloidal system, interacting via a square-well
potential, defined as in eq. (\ref{potential}) with $\epsilon= 0.03$,
solved within Percus-Yevick closure relation and calculated by solving
the MCT equations (\ref{mctstat}).  The horizontal axis represents the
colloid volume fraction $\phi$ and the vertical axis the temperature,
in units of $K_B/u_0$. The curve labelled as $B_1$ represents the
fluid-repulsive glass transition which is in agreement with the
hard-sphere limit for MCT at large temperatures (vertical dashed line
at $\phi\simeq 0.516$). The curve $B_2$ represents the
fluid-attractive glass transition. In the inset, it is shown in more
detail the attractive-repulsive glass transition curve with its
end-point, labelled as $A_3$. Also, the dashed curve in the low volume
fraction region represents the gas-liquid spinodal. 
}
\label{fig1}
\end{figure}

To discuss our results we shall frequently refer to the phase diagram
in Fig.  \ref{fig1}.

Let us begin by looking at what is at first sight one of the more
striking predictions of the calculations, the transition  between two types
of glasses and their merging at an end-point that has been labelled as
the $A_3$ point \cite{gotze}, since it represents a third order
singularity of the MCT equations (\ref{mctstat}).
 
The reader should be aware that,  
within MCT, on the transition lines, whether they be liquid-glass or
glass-glass, the two states of matter do not have any difference in
density or in structure and are differentiated in terms of the
non-ergodicity factors $f_q$.  Since this is a non-equilibrium
property of the system, we note that there is essentially no
equilibrium quantity that establishes the difference in phases, and
the order parameter must therefore be composed of $f_q$.

Since the repulsive glass and the attractive glass illustrated in
Fig. \ref{fig1}, and inset to that figure, are differentiated only by
the changes of their $f_q$ it is therefore natural to ask what
differences are implied in the shear modulus and other mechanical
properties of these two glasses, and how these difference vanish as we
approach the end-point where the two glasses become identical. We have
examined this question for the example of the shear modulus.

\begin{figure}
\centerline {\epsfxsize=17.0cm
             \epsfbox{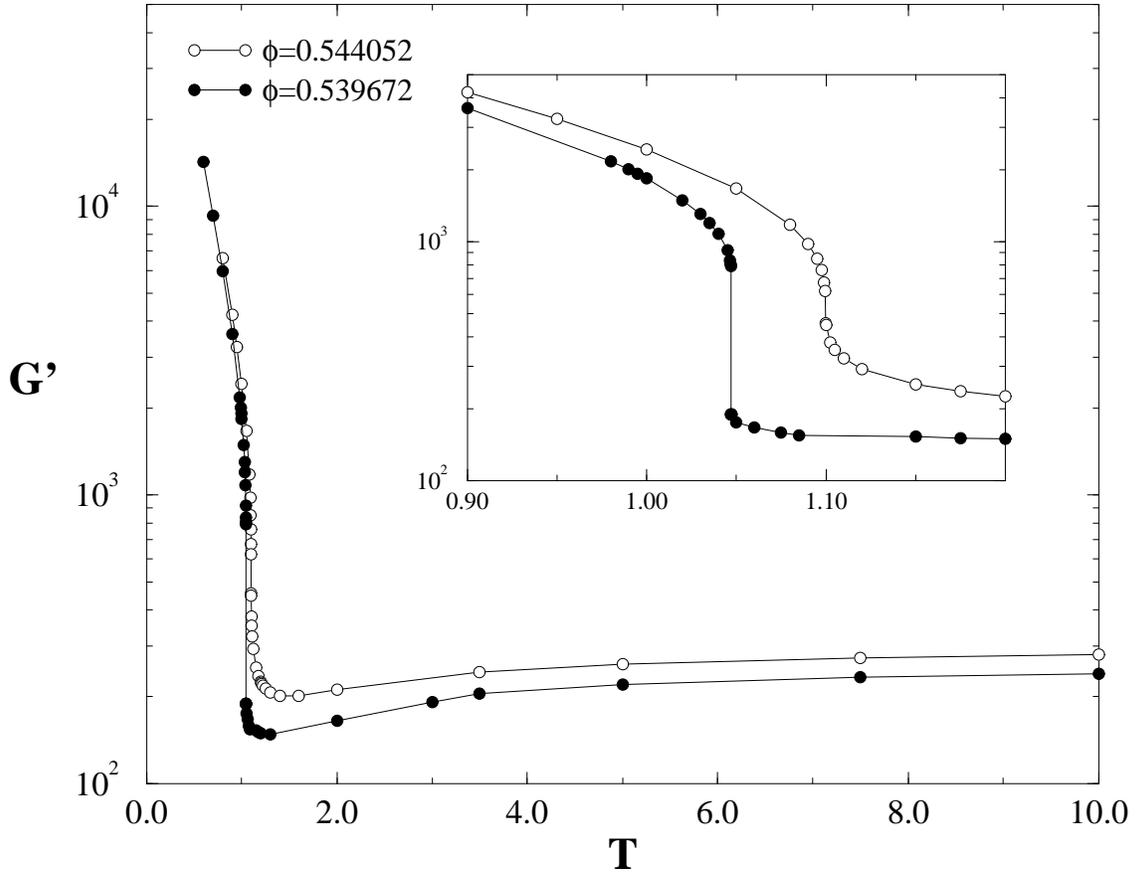}}  
\caption{
Plot of the elastic shear modulus $G'$ in function of temperature, for
$\phi=0.539672$ (black circles) and $\phi=0.544052$ (white
circles). The first value corresponding to crossing the glass-glass
critical line, while the second corresponds nearly to the endpoint
$A_3$ of this line.
}
\label{fig2}
\end{figure}

Thus, in Figure \ref{fig2} we plot $G'$ as a function of increasing
temperature for two fixed volume fractions $\phi=0.539672$ and
$\phi=0.544052$, both of them involving a crossing of the glass-glass
transition curve. For the smaller of these two volume fractions, it is
evident from the figure that, upon crossing the transition, there
appears a sharp discontinuity in the shear modulus. For the larger
one, that is very close to the end-point value packing fraction,
$\phi_{A_3}$, crossing the curve there is negligible difference
between the shear moduli of the two glasses.  Evidently, it is of
interest to define $\Delta G'$, the difference in $G'$ found in the
two glasses at the transition, and to examine this as a function of
$\Delta \phi$, and $\Delta T$, respectively the differences in volume
fraction and temperature from their end-point values, which we
evaluated as $\phi_{A_3}\simeq 0.5441$ and $T_{A_3}\simeq 1.09975$.
We numerically find, over the whole range of the glass-glass
transition, that the laws connecting these two quantities are,
\begin{eqnarray}
               \Delta G'\sim (\Delta \phi)^p; & \Delta \phi=
               \phi-\phi_{A_3} \\ \Delta G'\sim (\Delta T)^q ; &
               \Delta T= T-T_{A_3}
\end{eqnarray}
where $p$ is approximately $(0.33\pm0.03)$ and $q$ approximately
$(0.32\pm0.08)$. The errors in these estimates could be reduced with
larger computational effort, but there do indeed appear to be power
laws to high precision.  These exponents can be explained by a simple
argument.  We know that near an $A_2$ singularity $f_q$ is a solution
of a second order polynomial, while near an $A_3$ point it is solution
of a cubic one.  This implies,
\begin{eqnarray}
\label{expa}
f_q-f_q^{A_2} \sim (\epsilon_{A_2})^{1/2} \\
f_q-f_q^{A_3} \sim (\epsilon_{A_3})^{1/3}
\end{eqnarray}
 where $\epsilon_{A_i}$ can be both $T-T_{A_i}$ or $\phi-\phi_{A_i}$.
 Thus, evaluating the expression for $G'$ (\ref{G'}) in the proximity
 of the singularity $A_3$, this gives, in the leading order, an
 exponent $1/3$.

It will be of interest for experimentalists to seek the proposed
phenomenon of glass-glass transition, and to deduce estimates of any
exponents that arise.  Amongst the most accessible of these is that of
the shear modulus discussed above.  However, we do note the
reservation that it is likely that there is some associated structural
and density relaxation at the glass-glass transition, and this may
disturb the simple power law outlined above.

Evidently, a fairly practical comment that emerges from these results
is that the repulsive glass stiffness with respect to shear is much
smaller than the attractive glass one, as indicated by the large
vertical discontinuity shown in the inset of Fig. \ref{fig1} . This
reflects the fact that particles in the attractive glass are bonded by
the stickiness of the potential, whilst in the repulsive one there is
no real bonding between them, thus implying that they are more easily
broken apart under shear. Also, as expected, the attractive glass
shear modulus increases considerably with the decrease of temperature,
the attractions between particles becoming more relevant, whilst for
the repulsive glass there are no significant changes with temperature,
there being no energy scale involved in its formation.

Having outlined our results for the shear modulus differences between
the two glasses, it is perhaps worth while to revisit the physical
meaning and implications of these results. Firstly, all of the
differences in mechanical properties here come from the non-ergodicity
factor alone, rather than static structure since the two states have
the same structure factor, that is essentially the structure factor of
the liquid to which the MCT glass is referred. It is therefore
interesting to note that the differences in $f_q$'s between the two
types of glass, reflecting as they do the nature and efficiency of the
relaxation processes at different length scales, lead to such large
differences in moduli.
 
Let us now turn to another part of the phase diagram of these systems.
We have earlier alluded to the work by Verduin and Dhont
\cite{verduin} for a very short-ranged attractive colloidal system,
where they found some non-ergodic states at low temperature. These
they interpreted as gel states, but in line with the ideas laid out in
this paper they might also be viewed as attractive glasses, as it was
already discussed in \cite{fuchs}.  In fact, the horizontal portion of
the glass-liquid curve in their phase-diagram already is indicative
that the energy scale is playing a leading role, rather than packing
forces.  Therefore, we believe that it will transpire that it is quite
feasible to prepare and study attractive glasses in some detail.

In particular it will be possible to study many properties along the
liquid-glass transition. We comment first that any attempts to design
colloidal particles with the appropriate well shape will most likely
involve some uncertainties. Thus, the obvious methods of coating the
spheres with some attractive layer, or using depletion forces in
polymer solutions, cannot hope to exactly reproduce the parameters and
shape of the square well.  Therefore, the curve of liquid-glass
transitions corresponding to the square well, may be fit to
experimental data. Though it may not be correct in all details, it
should be reasonably accurate in its behaviour as a function of
packing fraction of spheres, and in reproducing many of the attendant
phenomena.  Thus, irrespective of the extent of agreement between the
square well potential and the effective potential that is ultimately
tested in experiments, we may be fairly certain that some predictions
of our theory will be more robust than others. Amongst these we might
again include the typical evolution of the mechanical properties as a
function of packing fraction.

\begin{figure}
\centerline {\epsfxsize=17.0cm
             \epsfbox{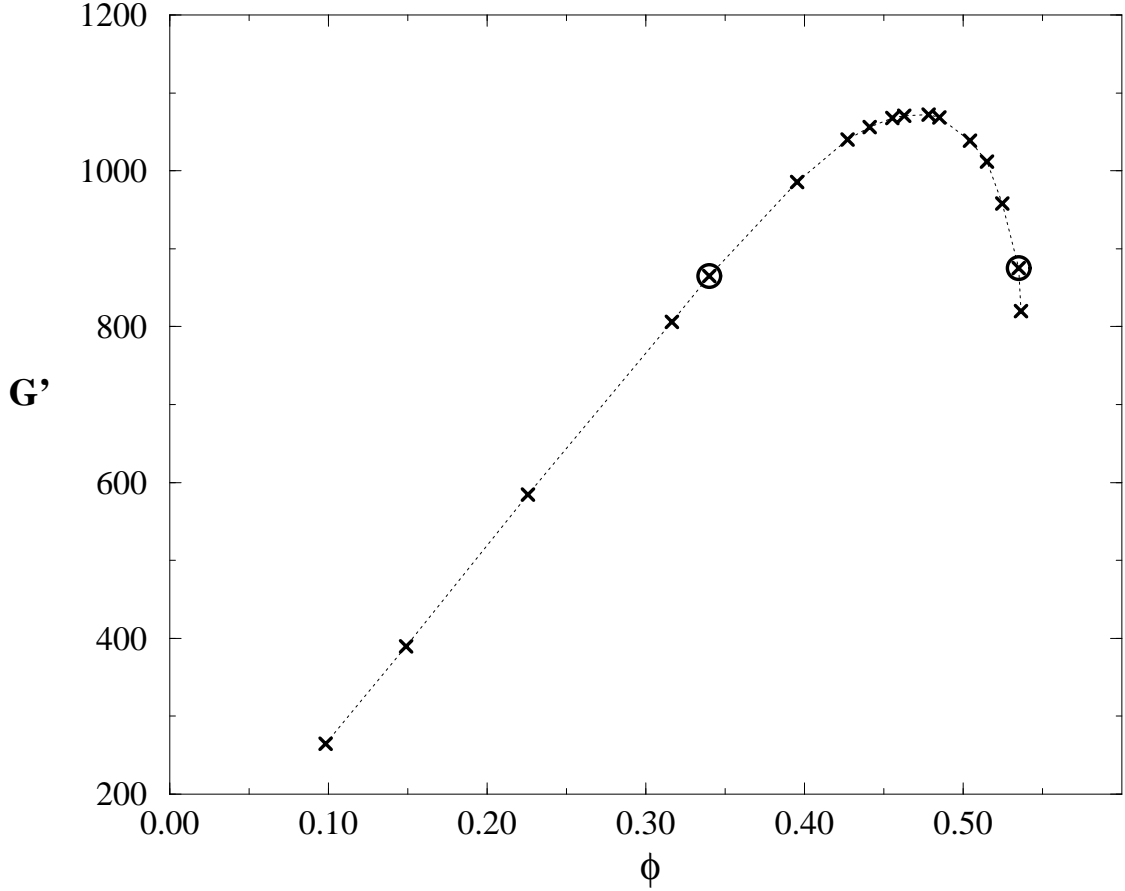}}  
\caption{
Plot of the elastic shear modulus $G'$ of an attractive glass as a
function of the volume fraction $\phi$, along the liquid-glass
transition, corresponding to packing fraction values up to $\phi
\simeq 0.535$ (as in Fig. 1). We note that the predicted value of the
volume fraction for hard spheres to undergo a glass transition is
$\phi_{HS} \simeq 0.52$. However, the decrease of the magnitude of the
shear modulus starts before we reach this value, at approximately
$\phi \simeq 0.48$. This corresponds to the stiffest glass we can
produce along the transition curve.
}
\label{fig3}
\end{figure}

Therefore it becomes of some interest to construct the evolution of
$G'$ on the glass side of the liquid-glass transition across an
extensive range of packing fractions encompassing all of what would be
viewed as the attractive glass, to be able to compare these results
with experiments.  Thus, in Figure \ref{fig3} we present the curve of
$G'$ as a function of $\phi$, along the attractive glass line,
labelled $B_2$ in Fig. \ref{fig1}.  We note that this curve extends
to quite low volume fractions, and that it is linear in a large range
of volume fractions. In fact this is true for all those values of
$\phi$ where the attractive interactions as considered to be
completely dominant. Ultimately the curve turns downwards towards
typical repulsive glass values of $G'$ when we go to higher volume
fraction.  Even so, we might have expected that the downward bending
of the curve would have happened very close to the critical hard
sphere packing fraction, $\phi \simeq 0.52$, whereas it commences
around $\phi \simeq 0.485$. Very similar characteristics for the shear
modulus, along the attractive glass transition line, have also been
found for the attractive Yukawa potential \cite{fuchs}.

\begin{figure}
\centerline{\epsfxsize=17.0cm
            \epsfbox{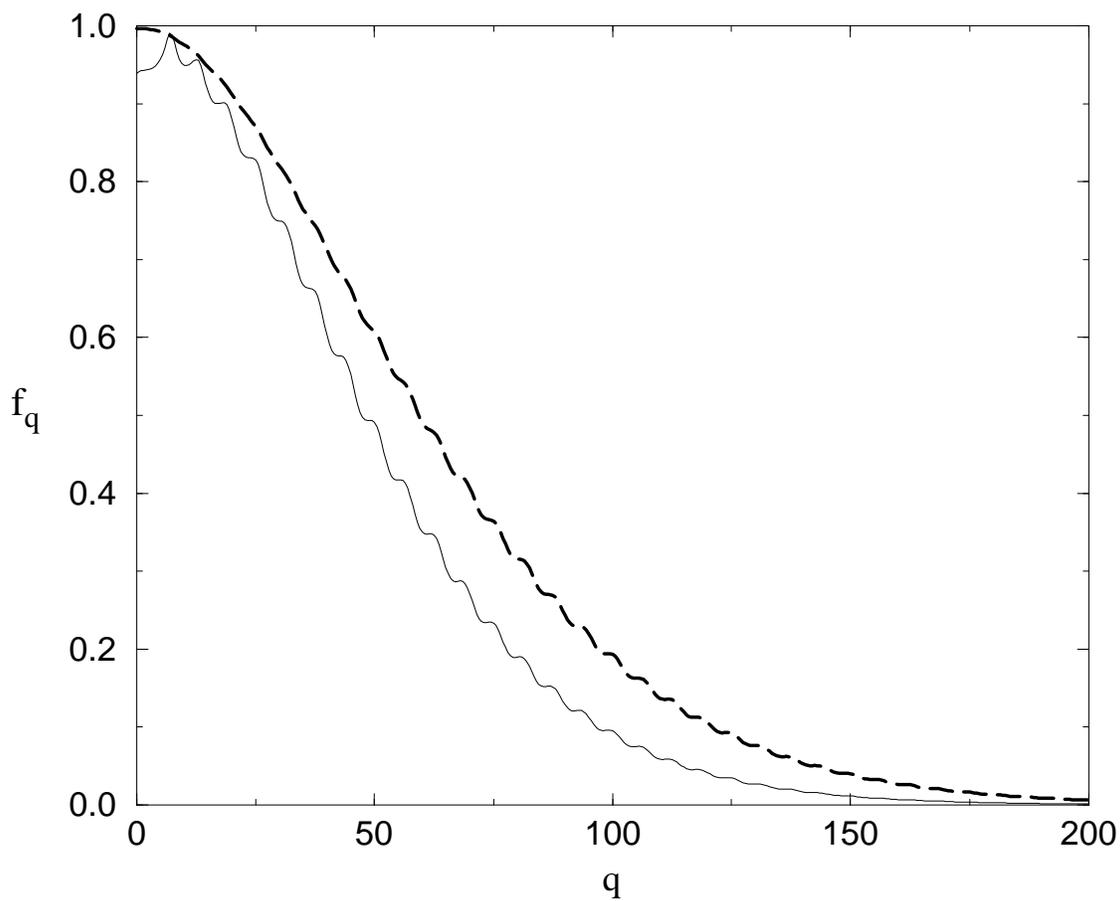}}  
\caption{
Non-ergodicity factors at the glass transition, corresponding to the
circled points in Fig. 3, where $q$ is expressed in units of particle
radius ($r=1$). The full line represents the non-ergodicity factor on
the repulsive side, and the long dashed line represents the state on
the attractive side. The non-ergodicity factors are certainly
different, but it is of interest to consider Figure 6 where we learn
that these difference in non-ergodicity factor alone is sufficient to
cause the considerable softening of the glass as the volume fraction
exceeds its critical value of $\phi \simeq 0.48$.  }
\label{fig4}
\end{figure}

Finally as a matter of curiosity let us draw attention to one aspect
of the shear modulus plot in Figure \ref{fig3}.  Consider the two
circled points on the curve in Fig. \ref{fig3}. They correspond to
states having approximately the same shear modulus but quite different
packing fractions, i.e. $\phi_1 \simeq 0.34$ and $\phi_2 \simeq
0.535$. Thus, we represent in Fig. \ref{fig4} the respective
non-ergodicity parameters of the two states for comparison. It is
interesting to note that, whilst the range of the two is almost the
same as it should be since they both represent states of attractive
glass, the one correspondent to the lower packing fraction exhibits a
more pronounced non-ergodicity of the system at every length
scale. The fact that we have the same shear modulus is a reflection
that both the non-ergodicity and static structure factors are relevant
for the modulus, and in this case we see that they compensate each
other so that we can `build', from the same system, two glasses with
the same stiffness with respect to shear, but having completely
different packings, and different structure.  This would be an
interesting phenomenon, if confirmed by experiments.

Now it is worth considering  the MCT prediction that glasses can exist at
volume fractions less than $10$\%, as indicated in the phase diagram, Fig.
\ref{fig1}.
Certainly attractive glasses can and should form at much lower
fractions than repulsive glasses, but we must not accept the results
of MCT blindly.  Note carefully the limitation that MCT does not
permit the self-consistent relaxation of structure in the glass phase,
and consequently the density cannot adapt to the new situation as the
ergodicity is lost.  The consequences of this have been outlined for
the glass-glass transition, but one should be aware also for the
possibility for qualitatively mistaken predictions in the region of a
liquid-glass transition where there is a meta-stable liquid-gas phase
separation, as in Fig. \ref{fig1}.  Thus we must acknowledge the
tendency for separation into more and less dense phases that could be
superimposed on the liquid-glass transition curve $B_1$. That this
does occur is not proven, but if it did, our MCT calculation would not
accommodate it.

Given this concern, we have calculated the mean number of bonded
spheres $n_b$ to a central sphere at the formation of the attractive
glass, along the curve $B_2$. We consider as sufficient condition for
bonding that the distance between two particles is up to the
attractive well width.  Thus, it is easy to estimate $n_b$ by carrying
out the integration over the pair correlation function such that only
those spheres falling within the attractive well are included in the
integration.  From the definition of the potential (\ref{potential}),
we then have
\begin{equation}
\label{nbond}
n_b=\int_0^{(d+\Delta)}\rho g(r)d{\bf r}
\end{equation}

For convenience we also calculated one estimate of the mean
coordination number $n_c$ around a sphere, irrespective of whether the
spheres are bonded or not.

 Nevertheless we have,
\begin{equation}
\label{ncoord}
n_c=\int_0^{r^*}\rho g(r)d{\bf r}
\end{equation}
where $r^*$ corresponds to our estimation of the first peak of the
pair correlation function $g(r)$.
 
The mean bonding number is well defined quantity, and is calculated
exactly here, whereas the mean co-ordination number is not so well
defined. The former may be relied on, the latter is used only as a
comparison to the mean bonding number.

\begin{figure}
\centerline {\epsfxsize=17.0cm
             \epsfbox{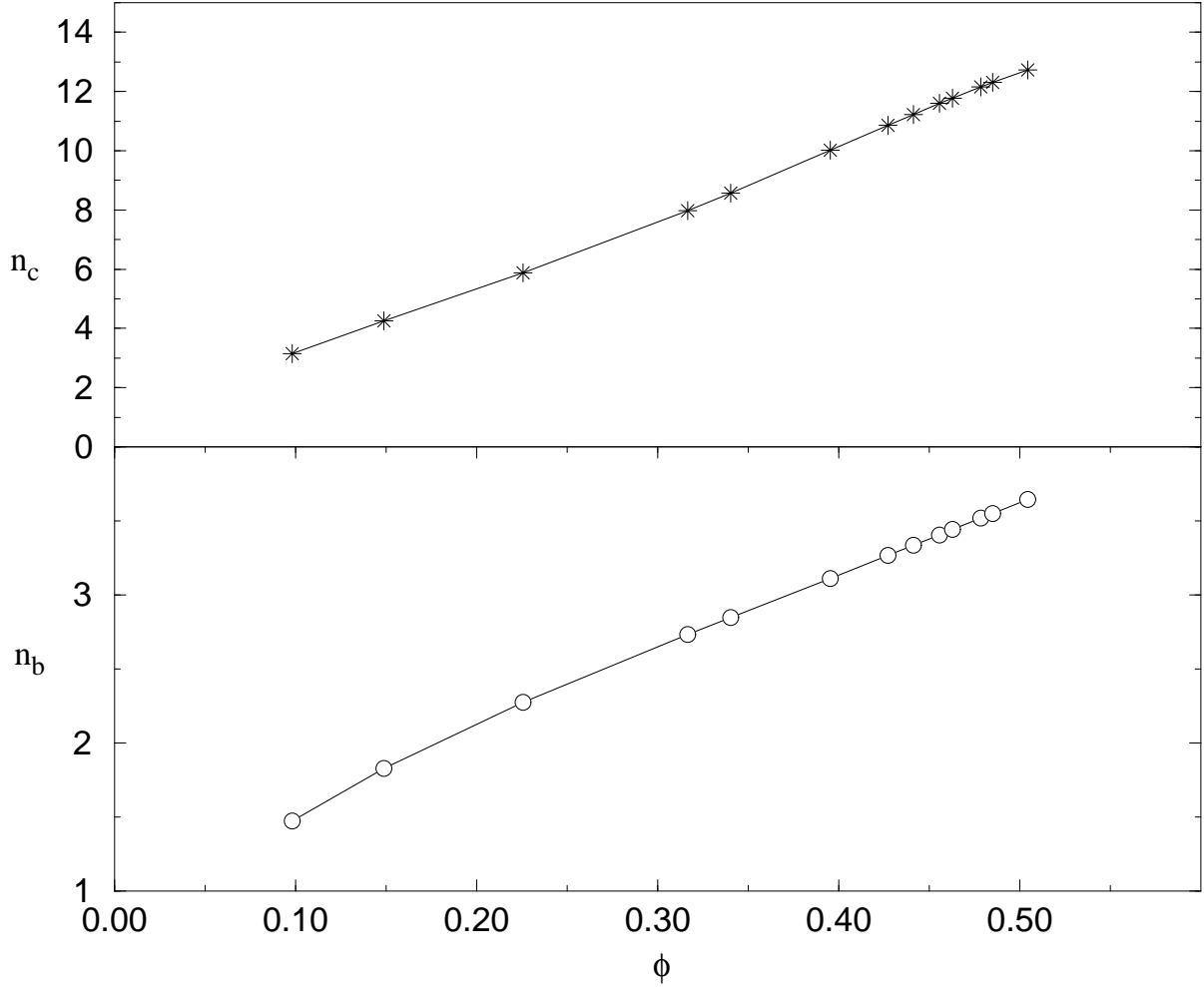}}  
\caption{
Plots of the coordination number $n_c$ (\ref{ncoord}) and of the mean
bonding number $n_b$ (\ref{nbond}) as a function of the volume
fraction $\phi$, along the liquid-attractive glass transition line
$B_2$. The two behaviours are nearly linear, except for small
deviations, at low volume fraction for $n_b$. As discussed in the
text, in this region, where $n_b$ is less than $2$, we question the
existence of the attractive glass, as predicted by MCT, because there
are not enough bonds between the particles to allow the formation of a
rigid structure.
}
\label{fig5}
\end{figure}

Both results are plotted in Figure \ref{fig5} as a function of the
volume fraction, along the curve $B_2$. It is striking that both
curves are nearly linear. However, the interesting thing to note is
that the mean bonding number at $\phi\simeq 0.10$ is around
$1.5$. Also, we can roughly estimate that only at $\phi\simeq 0.17$ we
find $n_b \simeq 2$.  Thus, it seems unlikely that extended
mechanically stable structures can exist for such small bonding
numbers. Indeed, a mean bonding number of $2$ would imply polymeric
structures, and for bonding numbers a bit larger than $2$ we may have
enough cross-links to establish a network, and finite shear
moduli. Just how large the mean bonding number has to be before a
finite shear modulus is obtained, we cannot say. Very simple arguments
based on counting degrees of rotational and vibrational freedom, and
requiring that there be no zero modes of the energy, indicate that the
minimum number of bonds to form an extended structure for a chemical
glass, should be about $2.4$ \cite{cusack}.  However these arguments
have not been shown to be relevant to the present case, so there is
little more progress that can be made now.

The discussion should, however, alert the reader to the possibility
that somewhere along the $B_1$ line, possibly when the bonding number
is a little less than $2$, the reduction of any further bonds in the
system might cause it to decompose, presumably into a less dense
and a more dense phase. These remain open questions for the moment,
but as MCT becomes applied more to glasses driven by attractions, it
will be important to consider them in future.

Of course, another fundamental question that should be addressed in
future is the applicability of idealized MCT to colloidal systems with
attractive interactions. Indeed, whilst for hard-sphere type systems
the idealized theory has been found in good agreement with
experimental results, it is not yet clear whether for attractive
systems, at lower densities, activated processes may become important,
so that hopping should be included in the theory for a better
description of this type of systems. Our present opinion is that the
idealised theory is likely to be useful for moderate densities, where
the bonding number is somewhat larger than 2. We do not preclude the
possibility that, for much lower densities, the theory may still be
suggestive, and indeed there have been already attempts to interpret
these regimes \cite{fuchs,fuchs2,fuchs3}.

Thus, all of this discussion leads us to seek a somewhat deeper
understanding of exactly what the basis of the mode coupling
predictions are.  By this we mean, we seek to understand the essential
features of the theory as it relates to attractive glasses.  One of
the points that we would like to probe a little more is to understand
just how local MCT is in its understanding of the loss of ergodicity,
and in its estimates of the main features of mechanical moduli.  The
simplest possible proposition would be that, essentially with
knowledge of the number of bonds and strength of associations between
nearest neighbour particles, we would estimate a correct MCT
transition, and, perhaps, even estimate a correct mechanical
response. We now seek to test this hypothesis. We shall examine this
question in two steps, beginning with the calculation of the
non-ergodicity factor that determines the glass transition itself,
then progressing to the shear modulus.

For simplicity, we call $q^*$ the wave-vector corresponding to the
first peak of the structure factor.  We then proceeded recalculating
the liquid-glass transition using two approximations.  Firstly, we
used the correct short wavelengths behaviour of the structure factor,
for $(q>q^*)$, in solving the MCT equations (\ref{mctstat}), but
using, for the long wavelength part of the structure factor
(i.e. $q<q^*$), the correspondent part of it for a packing fraction
that we chose arbitrarily.  We both considered $\phi\simeq 0.31$ and
$\phi\simeq 0.39$, obtaining no significant changes in the results.
We found that the essential features of the trends and quantitative
evolution of the transition are set only by the short-ranged part of
the structure factor, then irrespectively of what we chose for the
long wavelength part of it.  Thus, length scales involving only the
hard core and attractive well of a spherical particle are sufficient
to determine the glass-liquid transition within MCT.  Indeed, we
investigated a few points along the curve $B_2$ and we also found that
the transition temperatures are reproduced within a few percents
error.

To confirm these results, we also tried to evaluate the curve $B_2$
using an opposite strategy.  We then used the correct long wavelength
part of the structure factor ($q<q^*$), and as short-ranged part the
correspondent for the same values of packing fractions as before. This
lead not only to much larger errors in the glass-liquid curve $B_{2}$,
but, what is more important, to the wrong evolution of that curve with
volume fraction, this meaning that the curve was not only shifted but
also changed in its shape.  We may therefore conclude that, whatever
the full content of MCT, one can estimate the location of the
transition curve of the attractive glass to high precision by knowing
only local information about the particles around a central particle.
Long-ranged effects therefore appear not to play a central role in
this aspect of the MCT description. The same observation had been made
earlier for the Yukawa model and for the square-well potential, both
solved within MSA \cite{fuchs,us}. Thus, this is a general feature of
the theory itself, independently on the model or the approximation
chosen to solve it.

We now turn to another natural question about the predictions of MCT.
After having analyzed what the dominant contributions are for the
prediction of the location of the glass transition, we want to
investigate what is important in the determination of the mechanical
properties of the glass within the theory. To do so, we still refer to
the results of the shear modulus $G'$ in Fig. \ref{fig3}.

Thus, we note that, in equation (\ref{G'}), there are evidently two
important contributions to the shear modulus, the non-ergodicity
factor, and the logarithmic derivative of the structure factor.  We
may ask if only one of these features provides the main contribution
to the shear modulus. To check this we may proceed as before, first
retaining the correct structure factor, and using the non-ergodicity
factor of a chosen value of packing fraction ($\phi=0.39$), and then
vice versa.  The question may be further refined by asking whether it
is possible to further locate the main driving force of the shear
modulus as being short ($q>q^*$) or long ($q<q^*$) length scales.

\begin{figure}
\centerline {\epsfxsize=16.0cm
             \epsfbox{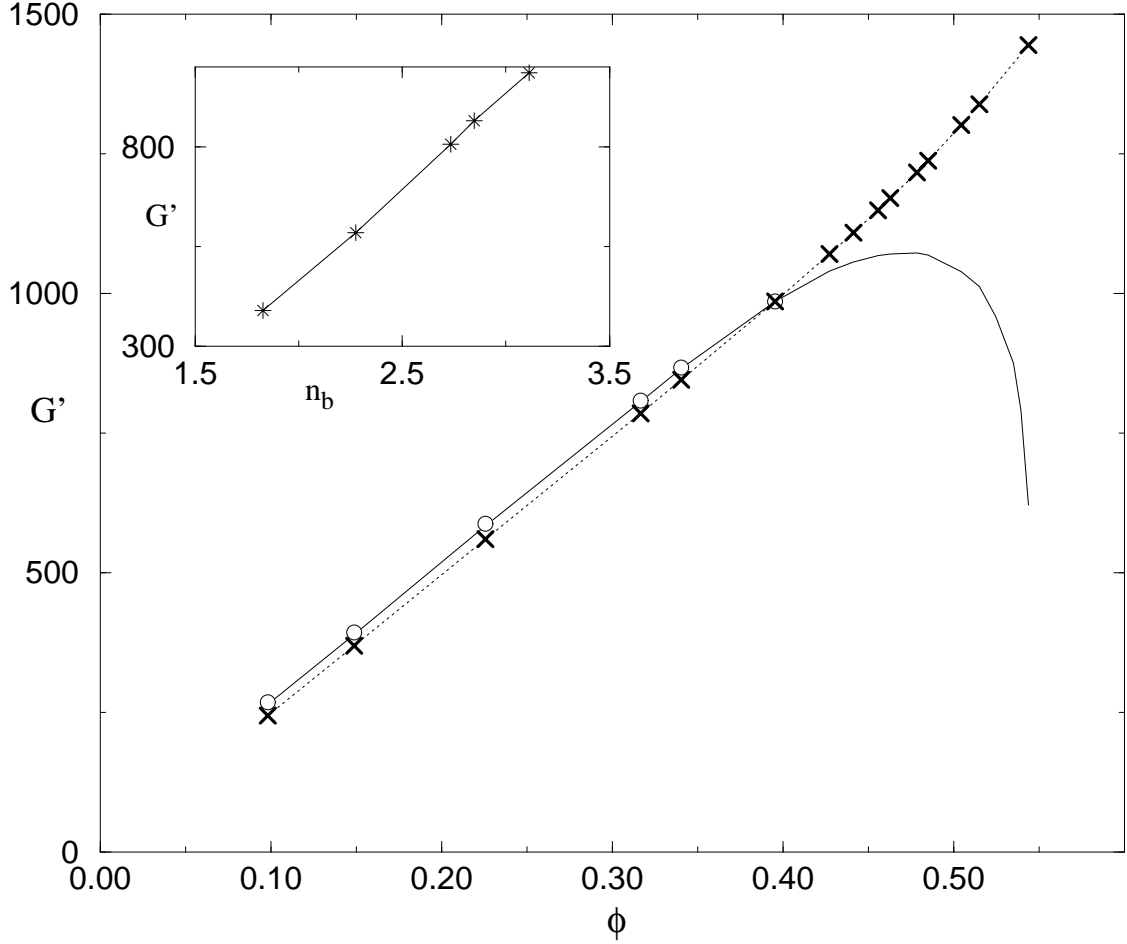}}
\caption{
Plot of the elastic shear modulus $G'$ as a function of $\phi$, as in
Fig.\ref{fig3} (full line). Superimposed we have plotted $G'$
calculated by using an incorrect reference $f_q$ (see text) and the
correct structure factor (crosses). Results obtained by using the
correct non-ergodicity factor and only the exact short length scale
contribution for $S(q)$ have been plotted using circles.
Together these approximate curves indicate that for most of the range
of stability of the attractive glasses its properties are dominated by
the short length-scale of the structure factor.  Also, providing the
system is non-ergodic ($f_q>0$), the degree to which it is so is not a
very important parameter in determining the mechanical properties we
discuss.
However, at higher densities the shear rigidity of the glass begins to
decrease again, as we approach a repulsive glass. This behaviour
cannot be described as above, and we now need to use the correct
non-ergodic factor, and the details of the structure factor now
becomes less relevant in determining shear rigidity.
Inset: Plot of the elastic shear modulus versus the number of bond
$n_b$, showing an almost linear dependence. Here we see that for the
true attractive glass the shear rigidity is close to linearly
dependent on the number of nearest neighbors within the range of
attraction.}
\label{fig6}
\end{figure}

The answer is striking.  With almost quantitative accuracy we find
that the linear behaviour of the shear modulus with volume fraction,
in the purely attractive region, which we can define being in the
range of volume fractions between $0.17$ and $0.4$, originates solely
in the short length-scale ($q>q^*$) behaviour of the structure factor,
as shown if Figure \ref{fig6}.  We represent here the curve, already
reported in Fig. \ref{fig3} , with superimposed the data obtained by
using the chosen $f_q$ and the true structure factor (crosses), and
those obtained by using the true non-ergodicity factor and only the
exact short length scale contribution for $S(q)$ (circles).  Thus,
providing the non-ergodicity factor is finite, and on the correct
scale, the shear modulus is relatively insensitive across the whole
attractive glass region to its details, being only shifted of a small
amount.  We note that in the figure all the sets of data are
coincident at $\phi=0.39$, since this is the chosen value of reference
for the different cases.  Since we find, as before, that only the
short length scale picture of the system is really involved in the
determination of $G'$ (circles), we can explain the linear behaviour
of the shear modulus with packing fraction in the attractive region,
being directly related to the number of bondings.  Thus, we represent
$G'$ in function of $n_b$ in the inset of Fig. \ref{fig6}, and we
show that there also exists a linear relation between them.

Finally, we turn to analyze what happens for the phenomenology of the
shear modulus at higher packing fractions, beyond what we believe is
the pure attractive region, where $G'$ undergoes a rather rapid
decline.  We have pointed out already that, along the glass-liquid
curve in Fig. \ref{fig3} this rapid decrease of the shear modulus
happens as we approach $\phi\simeq 0.48$, that is a value quite below
the close-packed structure appropriate for the hard core.  It may be
argued that this phenomenon is hardly surprising since we know that,
finally, the shear modulus must decrease to values characteristic of a
repulsive glass.  These are much smaller since, as we have earlier
pointed out, there are no attractive interactions, and no effectively
bonded particles to the central one.  However, it is nevertheless
interesting to understand why the shear modulus softens quite
dramatically at that particular packing fraction, and by what means
the effective attractions are being screened in the system in this
region of densities.

To understand this point, we have studied the thermodynamic pressure
for a square well potential, given in ref. \cite{croxton}.  In
particular, we have examined the pressure and compressibility of the
liquid along the glass-liquid curve, in order to exclude the
possibility that there be any anomalies in the liquid itself in the
relevant region.  We found no anomalies within the PY approximation in
the region of packings from approximately $0.45$ up to beyond the
endpoint value $\phi_{A_3}$.  It is then interesting to note that,
whilst the liquid is perfectly normal, the proximate glass, undergoes
this softening at well defined values of $\phi$ and $T$, corresponding
to the decrease of the shear modulus.  In essence we find that the
softening of the glass, whilst it certainly originates in
cancellations between hard core and attractive parts of the potential
as the density increases, occurs mechanistically within the MCT memory
kernel.

To be more precise, we have used the same strategy as in the purely
attractive region, to examine the different contributions of the
structure factor and of the non-ergodicity factor to the shear
modulus. So, in Fig. \ref{fig6} the data represented with crosses,
corresponding to the correct structure factor and a fixed arbitrary
non-ergodicity factor, are also represented at higher volume
fractions. It is evident that their linear behaviour, entirely due to
the structure factor, persists with increasing density. Conversely, by
a similar analysis, where the structure factor was chosen arbitrary
and the non-ergodicity factor was the correct one, we have found that
the principal origin of the decrease in the modulus is the change in
the non-ergodicity factor, since in this region it is changing from a
characteristic attractive to a repulsive form. The smaller
integral resulting in eq. (\ref{G'}) reflects the fact that the
attractive glass is less mobile due to the formation of attractive
bonds.  Thus, it can only be the non-ergodicity factor, solution of
the MCT equations (\ref{mctstat}), the responsible for the softening
of the glass.

\begin{figure}
\centerline{\epsfxsize=17.0cm
            \epsfbox{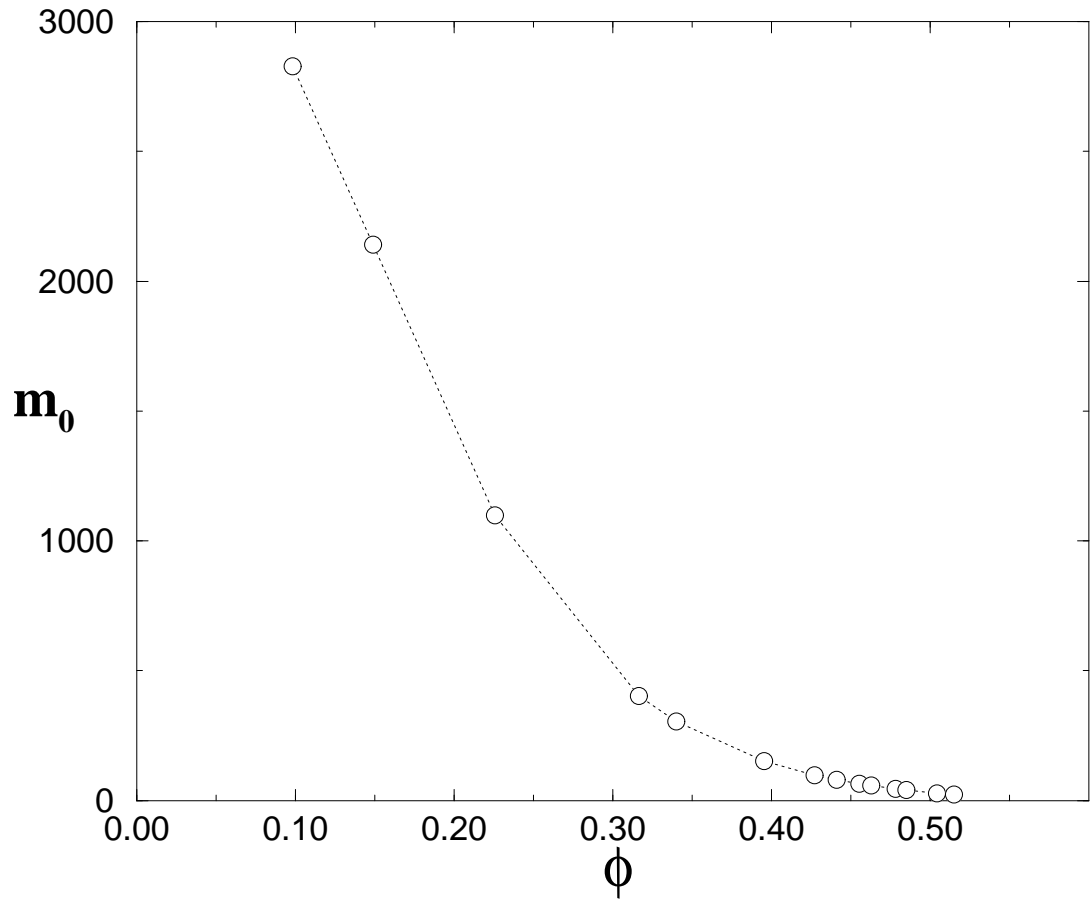}}
\caption{
Longitudinal stress modulus $m_0$ as a function of the packing
fraction $\phi$ along the part of the curve labelled $B_2$ in
Fig.\ref{fig1} corresponding to the transition between an attractive
glass and a liquid.}
\label{fig7}
\end{figure}

We now turn our attention to the behaviour of the longitudinal stress
modulus.  We have already shown that it presents a discontinuous
behaviour, similar to the elastic shear modulus, upon crossing the
glass-glass transition in \cite{us}.  Now we show $m_0$  along the
curve $B_2$, i.e. along the attractive glass-liquid transition line in
Figure \ref{fig7}, on the glass side of the transition, in the same
way as we did for the shear modulus. We conclude that, for increasing
packing fraction, though the glass becomes more rigid with respect to
shear, reaching a maximum at $\phi\simeq 0.48$, the longitudinal
modulus decreases continuously along that same curve, coming close to
its characteristic liquid value very close to the point where the two
curves $B_1$ and $B_2$ meet, corresponding approximately to $\phi
\simeq 0.536$. At first sight this seems counter-intuitive since we
expect that the extensional rigidity should increase with density, at
least up to that density where the elastic shear constantly increases.

\begin{figure}
\centerline {\epsfxsize=17.0cm
             \epsfbox{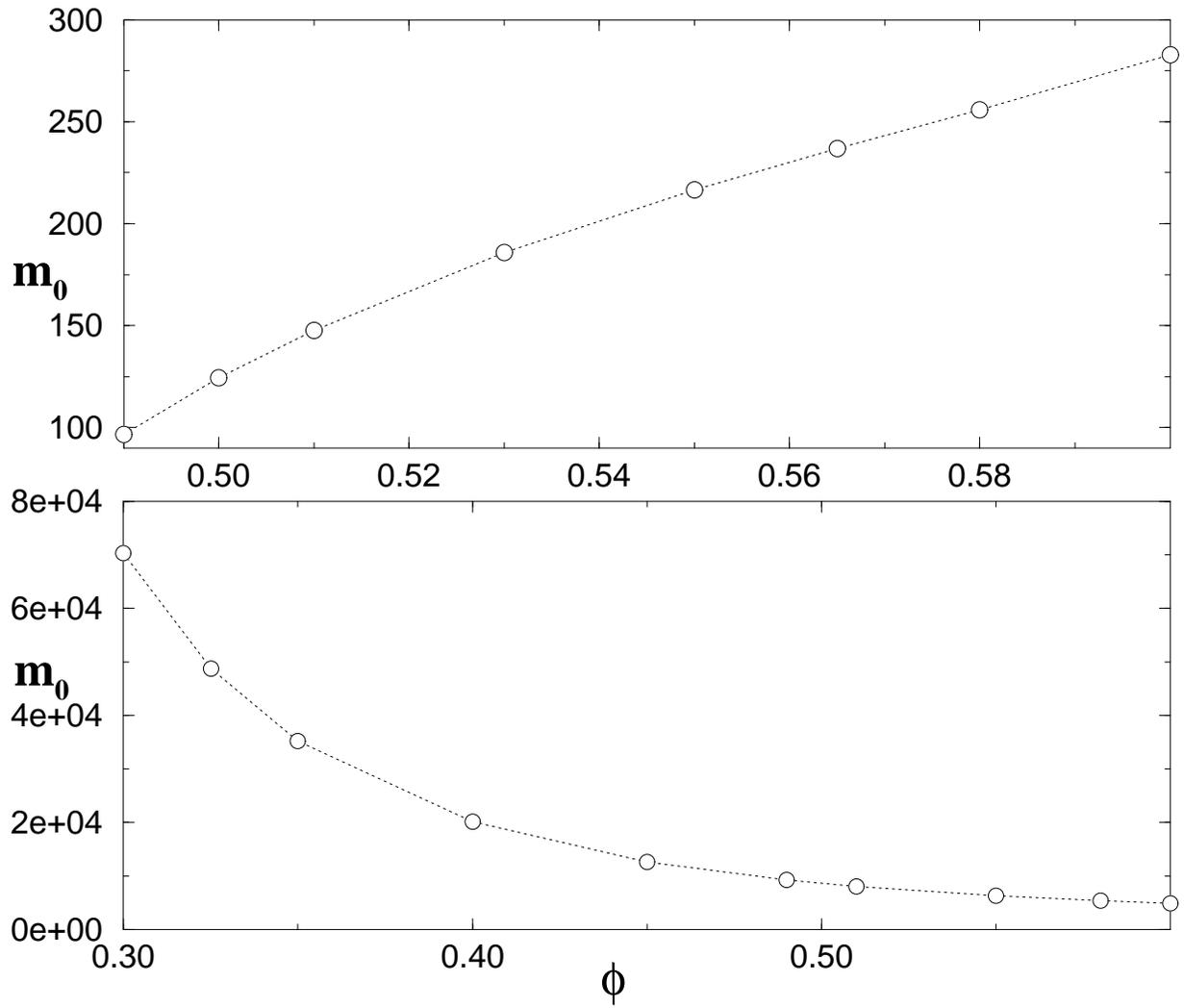}}

\caption{
a) Longitudinal stress modulus $m_0$ as a function of the packing
fraction along the isotherm $T=0.7$ .  
b) Same as graph a but along the isotherm $T=0.3$. 
 }
\label{fig8}
\end{figure}

If we examine the longitudinal modulus as a function of density, for
temperatures that cause the system to be within the repulsive glass
region only, the normal expectation of increased longitudinal modulus
with increasing density is confirmed, as illustrated in the $a$-part
of figure
\ref{fig8}.  In the $b$-part of fig. \ref{fig8}, instead, we have
the opposite behaviour of the longitudinal modulus with density for a
much lower temperature than in the $a$-part. Indeed, from
eq. (\ref{m0}), it is clear that the modulus depends on the
zero-momentum limit of the structure factor, appearing as a prefactor,
and the data correspond to a temperature for which the system  passes very 
closely to the critical point, or spinodal curve, of the underlying
liquid-gas transition.
In fact, MCT implies that the long length scale associated
with the proximate liquid, when it is near a critical point, or
otherwise the underlying spinodal, is quenched into the solid glass
because both liquid and glass have the same structure factor.  Within
MCT the large modulus in the attractive glass region therefore derives
essentially from this large quenched correlation length in the glass,
rather that any microscopic interaction.  We note that there is no
such quenched long length-scale dependence implied by the shear
modulus, either in the formula (\ref{G'}), or from the results showed
in Figs. \ref{fig3} and \ref{fig6}. In fact, as we showed earlier,
the shear modulus in the attractive region is mainly determined by
nearest neighbour adhesions.  The striking difference in the
behaviours of the two modulii is an interesting prediction of MCT,
when applied to attractive glasses. Possibly some of this variation
originates because long length scale structures are quenched into the
glass.  Whilst we do not know if it is true in nature, the strong
distinction between longitudinal modulus, and shear modulus along the
liquid-glass curve, will be readily testable in experiments on
colloidal systems.

\section{Conclusions}

We have studied the kinetic glass transitions of particles with a
model square well potential using Mode Coupling Theory.  This
interaction potential should, we believe, be a reasonable
approximation to that found in particles with grafted chains and
systems with strong depletion interactions. All of the non-trivial
results arise when the attractive piece of the interaction potential
is strong, but of very short range, and we have studied a typical
example of this type.  Based on this, we have been able to describe
some mechanical properties of two types of colloidal glass, the one
resulting mainly from attractive and the other mainly from repulsive
interactions between particles. We and others \cite{fuchs,sciotar,us}
have earlier proposed a distinction between the two kinds of glasses
based on their dynamical behaviour However, in this paper we have
shown that this difference may be probed experimentally using the
difference in mechanical properties, and that there are exponents that
describe the merging of these glasses into a single glass beyond some
critical density.

In general terms, glasses dominated by attractions have a stronger
rigidity under shear than those originating simply from packing
forces.

Also, we have studied the behaviour of the zero-frequency elastic
shear modulus $G'$ and the longitudinal stress modulus $m_0$ along the
liquid-attractive glass transition curve. Besides being of intrinsic
interest, this part of the study was chosen in the belief that
comparison to experiments will be more reliable along this curve, and
less dependent on model potential details.  The predictions are
striking. This is a positive feature since even confirmation of trends
will be of some interest in evaluating the MCT approach to these
systems.
 
At lower packing fraction, where we believe attractions to be
completely dominant, there is a linear increase of the shear modulus
with packing.  This reflects the fact that also the mean number of
bonds to a particle in the system is fairly linear in $\phi$ and the
only relevant physical mechanism to determine the shear modulus in
this region of the phase diagram occurs at short length scale.  We
have argued that here the shear modulus is determined by nearest
neighbour adhesions.  As an aside, we note also that within MCT the
attractive glass-liquid transition curve itself is essentially
determined by the the same factors as those determining the shear
modulus.

On the contrary, at higher values of packing fraction, the shear
modulus decreases quite dramatically towards typical repulsive glass
values. This phenomenon originates from the changes in the
non-ergodicity factor that compete in equation (\ref{G'}) with those
of the structure factor and lead the system to soften with respect to
shear.  The implication is that there should be a maximum of the shear
modulus along the liquid-glass transition.

Now we turn to the longitudinal modulus of the system.  Here the
result is quite different from that for the shear modulus.  Along the
glass-liquid transition curve $m_0$ decreases with increasing
density. We have observed that the transition curve, passes close to
the critical point and some portion of the spinodal of the underlying
liquid-gas transition, and consequently the input structure factor at
long wavelength is very large. The prefactor in formula (\ref{V0})
therefore becomes very large.  The physical implication is that large
scale density fluctuations quenched into the glass cause a large
increase in the longitudinal modulus, and the longitudinal modulus is
dominated by these for much of the attractive glass regime.

We note, in passing that MCT does not always reliably predict the
stability of the attractive glass. By this we mean that when we check
the mean bonding number to a particle in what has been predicted to be
a glass, we sometimes find that this number is even less than 2. We
conclude that in these cases, MCT is over-emphasizing the stability of
the glass, probably for a variety of reasons.  In any case, we use
this independent calculation as a rough check to exclude regions of
MCT glass that are clearly unphysical. It will be important to address
these issues in future.

There is no doubt that the model and the means by which we have
studied it are simple, and there are numerous limitations implied
thereby.  Nevertheless, the model is in some sense canonical in that
it contains the essential physical input of strong short-ranged
interaction, and repulsion. We believe that it encompasses many
essential ideas regarding the mechanical properties of colloidal
glasses and we have laid these out for consideration by further
theoretical, but mainly experimental researches.
 
We thank W. G\"otze and M. Fuchs for discussions and
suggestions. F.S. and P.T. are supported by INFM-PRA-HOP and MURST
PRIN98, E.Z., G.F. and K.D. by INCO-Copernicus Grant IC15CT96-0756 and
COST P1.

\end{document}